\documentclass[12pt,a4paper,preprint]{article}
\usepackage[latin1]{inputenc}
\usepackage{amsmath}
\usepackage{amsfonts}
\usepackage{amssymb}
\usepackage{graphicx}
\usepackage{bm}
\begin{document}
\title{Subleading Corrections to entropy formulae (convergences and
divergences)}
\author{Suhail Ahmad\footnote{corresponding author: suhail.dream@gmail.com} 
        , Sharf Alam\footnote{sharfalam@yahoo.co.in} \\
        Physics Department,\\
        Jamia Millia Islamia,\\
        New Delhi 110025, INDIA}
\date{\today}
\baselineskip 20.0pt
\maketitle
\begin{abstract}
\baselineskip 16.0pt
We know that sub-leading corrections to the hawking area law is riddled with issues which
have some convergent and divergent aspects. Depending on the theory, scheme, model or
even method sub- leading terms turn out to have trivial and non- trivial aspects which we are
going to dwell upon. The generic character of the first sub leading logarithmic term comes
out unanimously the same from all theories of quantum gravity like Strings, Loops, or even
semi-classical methods with the exception that sometimes the pre-factor of logarithmic term
turns out to be model dependent parameter or number hence consensus on this issue is
yet to be finalized. In this paper we will try to compare and contrast how we get the
corrections in various theories of quantum gravity including semi-classical methods on the
variant of Black Hole that is BTZ Black Hole .Towards the end we see how the addition of
chern-simon terms affects the entropy of black holes and we will make brief observations
regarding the same.
\end{abstract}
\section{Introduction}
We know from the standard result of  Bekenstein-Hawking result  that entropy is related with horizon area   as 
\begin{equation}
S=\frac{A}{4}+....
\end{equation}   
All theories of Quantum gravity should predict rather contain Bekenstein-Hawking entropy formulae to start with as such there is no surprise that all dominant approaches to quantum gravity whether strings, loops or semi-classical methods reproduce Bekenstein Hawking Formulae ab initio however, there are issues which dwell upon the fact that sub leading terms after taking quantum corrections into account can have convergent and divergent aspects.  
\section{Convergences and Divergences in Entropy Formula}
As we know that the quantum corrected formulae takes the form generally as following
\begin{equation}
S=A+b\ln (s)+ \mbox{const}+O(a....)
\end{equation}
Where b is logarithmic pre factor 
The common feature of all these theories is that after the first term, that is sub leading term the entropy is proportional to logarithmic of horizon area. The generic  sub leading term has a pre factor, unfortunately there seems to be  no consensus regarding the value of pre-factor. In the survey of literature the pre factor appears to be highly model dependent and method dependent parameter \cite{1}.When full quantum effects are taken into account the area law should undergo corrections and these corrections can be obtained by field theory methods quantum geometry techniques, general statistical arguments and most persuasively from Cardy formulae. Here we are going to take a variant of black hole BTZ Black hole and study sub leading corrections in different formalisms although the existing computations of black hole entropy have different starting points they share some common points which we shall review. Much of the interest in string theory regarding entropy formulae derivation was sparked by strominger vafa for extremal black holes \cite{2} where one could compute hawking entropy
By counting D brane states where the leading order correction turns out to be
\begin{equation}
S = \frac{A}{4}\mp \frac{3}{2}\ln{\left(\frac{A}{4G}\right)}\pm 2\ln Q + const
\end{equation}
For BTZ  black hole stromingers derivation gives
\begin{equation}
S=\frac{2\pi r}{4G}-\frac{3}{2} \ln{\left(\frac{2\pi r}{G}\right)}-\frac{3}{2} \ln{kl} + const
\end{equation}           
Exactly we get the same sub leading term from the quantum geometry formalism the computation has been carried out by Koul and Majumdar for non rotating black holes in 3+1 dimensions \cite{3}. Interestingly both of these formalisms use techniques of 2d conformal field theory to compute the entropy at the intermediate level although they have different starting points. The use of 2d conformal field theory in particular Cardy formulae makes it easy to count  states and its use  may not be merely a useful trick as there are suggestions that conformal field theory really provides  a universal description of low energy black hole thermodynamics.     Work of carlip, \cite{4} Strominger among others to calculate horizon  entropy is strong evidence in favour of the idea, In case of string theory if the symmetries contain copy of virasoro algebra and the algebra of corresponding charges also form a virasoro algebra, then the central extension of this algebra can be used in cardy formula to calculate entropy of the horizon in black holes. In quantum geometry the theory of irreducible representations of simplest of CFT theories. SU(2) Wess-Zumino model, is crucial step to yield the quantum generalization of semi classical B-H entropy of black hole \cite{3}. The striking feature in these formalisms  is that conformal symmetry can be enough to determine entropy of the horizon without knowing underlying QFT lending support to the notion that there is effective description of  horizon degrees of freedom suggesting  that the effective theory of quantum gravity is 2d, CFT \cite{5}. The main virtue of Cardy formula or other CFT techniques is also its main weakness as we can count states without knowledge of full Quantum Gravity whose actual states remain disguised. Here we would like to make the comment that higher order corrections should give us information beyond the effective description given by  all the current models of quantum gravity. It remains to be seen whether entanglement should play a role in connecting to the sub leading corrections or higher derivative corrections as has been suggested by some authors. There is a pertinent comment by Martinec that general relativity must be considered as effective field  theory  \cite{6}, which cannot distinguish among different conformal field theory states. One of the successes of most of the black hole-CFT comparisons in the literature can be traced back to the matching of symmetries and anomalies, this gives a better understanding of the entropies even match at sub-leading level \cite{13}. When we analyse   sub leading corrections via semi-classical methods   we find a universal logarithmic term \cite{7}, but whose pre factor turns out to be different and also its value varies for the type of black hole we are interested in, for example in a particular variant of BTZ black hole \cite{8} studying corrections to the entropy of charged and rotating black holes beyond semi classical approximations. The entropy including the correction terms turn out to be 
\begin{equation}
S=\frac{\pi(r^{2}+a^{2})}{\hbar}+\pi \alpha_{1} \ln{(r^{2}+a^{2})} +\sum_{k>2}\frac{\pi \alpha_{k-1} \hbar^{k-2}}{(2-k)(r^{2}+a^{2})^{k-2}}+..... 
\end{equation}                                                     
The above formula is arrived at when the exactness of differential of entropy on the three parameters of mass charge and angular momentum are taken at quasi-equilibrium states and then applying quantum corrections to laws of black hole thermodynamics \cite{9}. It is not strange that we should be able to recover hawking area result in much easier fashion while the issues of sign and numerical value of pre factors may not be the same as we get in other theories of quantum gravity. To add to the above discussion it is startling to know that from purely analytical number theoretical methods that Rademacher expansion beyond leading order exactly verifies the leading pre factor \cite{10} thereby giving us some seemingly convincing reason for veracity of sub leading term as something genuine to be accounted for. Thus to conclude the above discussion we may say that generic nature of sub leading term at leading order terms gives us impression that at first order entropy or area law is verified from diverse approaches that is all roads lead to the same place, However variation in sub leading terms may be a trivial or non-trivial effect which should give us veracity of any particular theory or formalism from which they are obtained from the first principles. We feel or it may be suggested also, that sub leading terms should give us information beyond effective degrees of freedom in gravity theories.
\section{Role of Chern-Simon terms in the entropy formula in BTZ black holes:}
The gravitational Chern-Simons corrected/dressed BTZ black holes has been studied in the context of higher derivative gravities \cite{11,12}. It is seen  that with the addition of chern simon terms lot of non trivial interesting stuff takes place in particular as we know due to the presence of two horizons  outer and the internal one  in BTZ black holes in the presence of chern simon terms role of entropy dependence turns out to be
\begin{equation}
S=\left(\frac{2\pi R_{+}}{4Gh}\right) + \beta \left(\frac{2\pi R_{-}}{4Gh}\right)
\end{equation}
  This has important bearing on how to preserve the second law in general when the chern simon coupling dominates and when its seen that entropy dependence is more on the inner horizon. So understanding the roles of inner horizon data appearing in the thermodynamics   relations would be challenging problem.On the general efficacy of Cardy formula with the higher derivative/curvature corrections it is interesting to know that statistical entropy from the cardy formula has basically the same form for both the Einstein-Hilbert action and gravitational chern-simons corrected action .
\subsection*{Acknowledgement}
S. Ahmad highly acknowledges the financial support from UGC India.


\begin{thebibliography}{99}
\bibitem{1} Medved A J M 2005 Class. Quant. grav. 22 133.
\bibitem{2} Strominger Vafa 1996 Phys. Let. B 379 99. 
\bibitem{3} Koul R K and Majumdar P 1998  Phys. Lett. B 439 267.
\bibitem{4} Carlip S 2000 Class. Quant. grav. 17 4175.
\bibitem{5} Hyeyoun C 2010 Dyanamics of diffeomorphic degrees of freedom at horizon. arXive:gr-qc/1011 0623.
\bibitem{6} Carlip S conformal field theory 2+1 dimensional gravity and the BTZ black hole gr-qc/0503022.
\bibitem{7} Saurya D Majumdar P Bhaduri R K 2002 Class. Quant. Grav. 19 2355
\bibitem{8} Akbar M and Saifullah K 2011 Gen. Relativ.  Gravit. 43 933.
\bibitem{9} Banerjee R and Modak S K 2009 J High Energy Phy. 5 63 
\bibitem{10} D.Bimingham ,I Sach, and S Sen, 1998 Phys. Lett. B 424 275.
\bibitem{11} Sahoo and Sen 2006 J High Energy Phy. jhep07 (2006) 008.
\bibitem{12} Park M BTZ black hole with gravitational chern simons; thermodynamics and statistical  entropy arXive:hep-th/0608165.
\bibitem{13}Lectures on AdS$_{3}$/CFT$_{2}$ Correspondence Per-Krauss arXive:hep-th/0609074v2
\end{thebibliography}
\end{document}